\documentclass[12pt]{article}
\usepackage{graphicx}
\usepackage{setspace}
 \usepackage{xcolor}

\begin{document}

\title{Imperfect bifurcations in opinion dynamics under external fields} 

\author{Francisco Freitas, Allan R. Vieira and Celia Anteneodo}

\maketitle

%\ead{celia.fis@puc-rio.br}
%\vspace{10pt}
%\begin{indented}
%\item[] September 2019	
%\end{indented}

\begin{abstract}

We investigate, through a kinetic-exchange model,  the impact that an external field,  
like advertising and propaganda, has on opinion dynamics. 
We address the situations where two opposite alternatives can be selected 
but the possibility of indecision also exists. 
In this model, individuals influence each other through pairwise interactions, which can be of 
agreement or disagreement, and there are also external fields that can skew decision making. 
Two parameters are used to model the interactions with the field:  
one measures the sensitivity of the individuals to be influenced, another 
quantifies in which direction. 
We study this model in a fully connected social network scenario,  
by means of numerical simulations of the kinetic exchange dynamics 
and analytical results derived from the mean-field rate equations. 
We show how the external bias gives rise to imperfect bifurcations, 
and cusp catastrophes, allowing abrupt changes and hysteresis 
depending on the level of disagreement in interpersonal interactions and on 
the strength of the external influence. 
\end{abstract}
\vspace{2pc}

\onehalfspacing

\section{Introduction}

Many models of opinion dynamics, mainly those coming from the statistical physicist's community, 
address self-organization aspects~\cite{castellano2009statistical}. 
In these cases, the purpose is to unveil how the interpersonal interactions, by themselves,  
contribute to shape public opinion.  
Moreover, assessing the conditions for the emergence of self-organized structures is very important {\it per se} since it helps to understand mechanisms that can be present in other complex systems. 
Despite in a social group collective states, like consensus or polarization, 
can emerge from the interactions amongst individuals, without the need of any external control~\cite{ramos2015does}, advertising and  propaganda are always present, 
taking advantage of a variety of media,  from conventional (such a television, radio, newspapers, outdoors, etc.)  
to  modern robots and influencers in the internet.
Therefore, their global impact and interplay with the internal interactions, 
which can promote the alignment to either one or other opinion~\cite{ferrara2016rise}, 
is another important social phenomenon to be studied.  
But, as far as we know, relatively few works in the vast literature on opinion dynamics address this issue. 
Among them let us mention a study, 
based on real data of mobile's oligopoly market, on how advertising 
influences the choice of a given operator~\cite{sznajd2008outflow}. 
In that case, a critical behavior depending on the level of advertising is observed. 
In another paper, the effect of mass media is investigated through a variant of 
Sznajd model~\cite{crokidakis2012effects}, where individuals follow 
an external field with certain probability. 
Shortening of relaxation times and suppression of the order-disorder phase transition 
for sufficiently strong field is observed, facilitating consensus, as expected. 
But many questions arise, for instance which is the interplay of contrarian and undecided 
people with the field, or which is the impact of opposite fields.

In the present  work, we focus on a binary scenario where people have two opposite choices 
(e.g., ``for'' or ``against'' a controversial issue, ``yes'' or ``no'' in a referendum, etc.) 
and indecision can also happen.  The inclusion of the undecided state is important since 
it is known to play a role in the route to 
consensus~\cite{lallouache2010opinion,biswas2011mean,biswas2012disorder,crokidakis2012role,lundberg2014decisions,balenzuela2015undecided,vazquez2004ultimate,svenkeson2015reaching}.
We consider that the individual opinions evolve due to interactions among people 
that exchange their thoughts and influence each other, both positively or negatively, 
and also due to external voices driving the dynamics towards one of the alternatives. 
This is a scenario found in many real situations, 
where the influence of biased information may be crucial, mainly nowadays, 
with high ease for fast and widespread dissemination. 
We investigate this question by means of a model where opinions evolve through kinetic exchanges. 
The details of the model are defined in Sec. \ref{sec:model}. 
In Sec. \ref{sec:results}, we present the outcomes of simulations in fully connected networks, 
interpreted in terms of mean-field rate equations. 
Remarks about the results of this model are discussed in Sec. \ref{sec:remarks}.

\section{Model}
\label{sec:model}

We consider an opinion formation model based on kinetic exchange 
rules of interaction~\cite{sen2011phase, biswas2011mean,biswas2012disorder, crokidakis2012role,lallouache2010opinion}. 
Each individual $i$, in a population of size $N$, possesses an opinion $o_i \in \{-1,0,1\}$, 
which indicates whether $i$ has a positive or negative position 
about the topic under discussion (i.e., $o_i = \pm 1$), or an undecided (neutral) 
attitude ($o_i = 0$). 

Opinions are updated via Monte Carlo (MC) simulations. 
At each iteration, two agents are randomly sorted and their opinions are updated. 
Updating  is governed by kinetic exchange  and by the influence of an external 
contribution that promotes alignment to one of the alternatives, according to 
\begin{equation}\label{model}
	\begin{array}{r@{}l}
		o_i(t+1) &= o_i(t) + \mu_{ij} o_j(t) + \Phi_i(t), \\
    o_j(t+1) &= o_j(t) + \mu_{ji} o_i(t) + \Phi_j(t),
	\end{array}
\end{equation}
where $\Phi_i$ represents the contribution of the external field 
to mold the opinion of agent $i$ 
and $\mu_{ij}$ is the strength of the influence of agent $j$ over $i$.  
Moreover, this influence can be either of agreement (positive) or disagreement (negative). Negative interactions can be seen as a kind of contrarian 
behavior~\cite{galam2002minority,galam20072000}, 
which means that agent $i$ tends to adopt the opposite opinion of agent $j$ (counter-imitation). 
If Eq.~(\ref{model}) yields higher (lower) values than $1$ ($-1$), then the opinion is set to the 
corresponding extreme value $1$ ($-1$).

In the current setup, we consider that the coupling is an annealed random variable, controlled by the parameter $p$ 
according to

\begin{equation}\label{mu}
    \mu_{ij} =
	\left \{
	\begin{array}{cl}
		1,	&  \textnormal{with probability} \ 1-p, \\
		-1, &  \textnormal{with probability} \ p,
	\end{array}
	\right.
\end{equation}
where $p$ represents the average fraction of negative interactions, 
that is,  of individuals which follow the contrarian behavior. 
Notice that in general $\mu_{ij} \neq \mu_{ji}$, but this is irrelevant in the annealed version.

The external contribution acting in Eq. (\ref{model}) is also an annealed random variable, 
defined as
\begin{equation}\label{F}
\Phi_{i} =
\left \{
\begin{array}{cl}
1,  &     \\
0,  & \textnormal{with probability}   \\
-1, &   
\end{array}
\right.
%\left \{
\begin{array}{l}
\phi \omega, \\
1-\phi, \\
\phi(1-\omega).
\end{array}
%\right.
\end{equation}
The external influence can be turned off by setting $\phi = 0$. 
In this case, it is well-known that the system undergoes a non-equilibrium 
order-disorder phase transition~\cite{biswas2012disorder,crokidakis2012role},
at the critical value of the fraction of negative interactions $p_c=0.25$. 
When $0<\phi\le 1$, the population is exposed and it is sensitive to the external influence. 
Then, the parameter $\omega$ measures the relative contribution of two opposite advertisement sources,  
favorable to each available alternative. 
According to Eq.~(\ref{F}), the action of the external fields is symmetric around $\omega=0.5$, 
under the transformation $\omega \to 1-\omega$, $f_k \to f_{-k}$, 
where $f_k$ is the fraction of individuals with opinion $k$. 
In particular, if $\omega=0.5$, both external fields contribute in the same measure, and the system is perfectly symmetric with respect to the two opposite opinions.
In the extreme cases $\omega=0$ and $\omega=1$,  a single external influence is present.
Due to this symmetry, we choose to analyze one of the intervals, namely $\omega \in [0.5,1]$, 
which corresponds to fields favoring opinion $o=1$. 

\section{Results}
\label{sec:results}
We run the Monte Carlo dynamics ruled by the algorithm defined 
by Eqs. (\ref{model})-(\ref{F}), in fully connected networks.
In order to followed the evolution of the collective state of the system, 
we measure the fractions $f_k$ of individuals with opinion $k=-1,0,1$.

\begin{figure}[b!]
\includegraphics[scale=0.9]{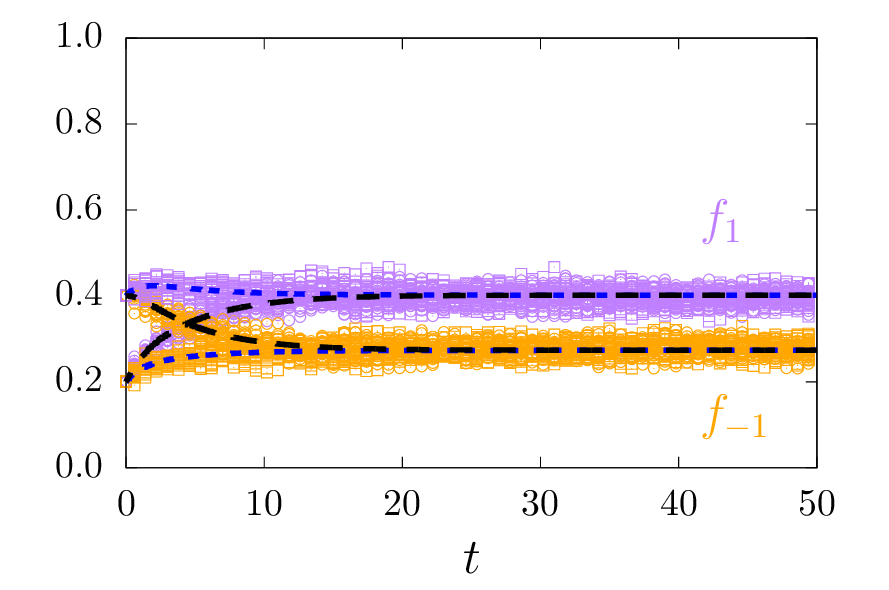}
\includegraphics[scale=0.9]{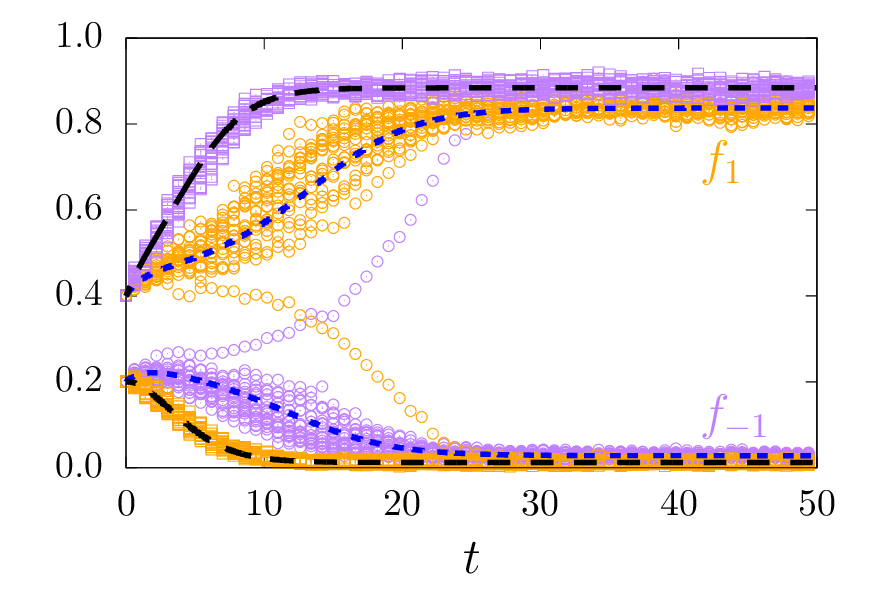} \\
\mbox{}\hspace{3cm}	  (a) 	\hspace{7.5cm} (b) 
	\caption{Time evolution of fractions $f_{1}$ (lilac) and $f_{-1}$ (orange), 
	for $\omega=1$, $\phi=0.05$, 
	and disagreement level: $p=0.4$ (left) and  $p = 0.1$ (right). 
	In each case, two initial conditions are considered: 
	$(f_{1},f_{-1})(t=0)$ = (0.4, 0.2) (circles) 
	and (0.2, 0.4) (squares), sorting 30 different configurations 
	in a network of size $N=1000$.	Dashed lines represent the mean-field calculation. 
	}
	\label{fig:evolution}
\end{figure}

In Fig.~\ref{fig:evolution}, we illustrate the temporal 
evolution of the fractions of both opinions, in 
the case $\phi=0.05$ and $\omega = 1$ (recall, that, by Eq.(\ref{F}), 
this setting favors option $o=1$), 
for two different levels of contrarian behavior. 
In both cases, two initial values of $(f_{1},f_{-1})$  are considered. 
We see that for the level of contrarians $p = 0.4$, in the left panel, 
independently of the initial conditions, $(f_1,f_{-1})$ evolves towards a state 
where the opinion favored by the field wins, despite $\phi<<1$. 
Differently, for a low level of contrarians (e.g., $p = 0.1$, in the right panel),  
the final collective state depends on the initial condition and even the opinion favored by the field 
can end up losing (squares). 
This indicates the existence of multistability. 
Finite size fluctuations ($N=1000$ in the example),  
 can alter the evolution, making it proceed following the fate of phase space points in other basin, 
as observed for one of the trajectories in panel (b).

This portrait can be understood analytically. 
Considering that the dynamics occurs on the complete graph topology,  
we obtain  the mean-field equations for the transition rates of each opinion

\begin{eqnarray}\label{taxes1}
\frac{d f_1}{d t} &=& G_1(f_1,f_{-1}, p, \phi, \omega), 	\\ \label{taxes2}
\frac{d f_{-1}}{d t} &=& G_{-1}(f_1,f_{-1}, p, \phi, \omega), 	
\end{eqnarray}
where the functions $ G_1$ and $ G_{-1}$ and their  derivation is given in the Appendix.

\begin{figure}[h]   %width=6cm,height=6cm
\includegraphics[scale=0.3]{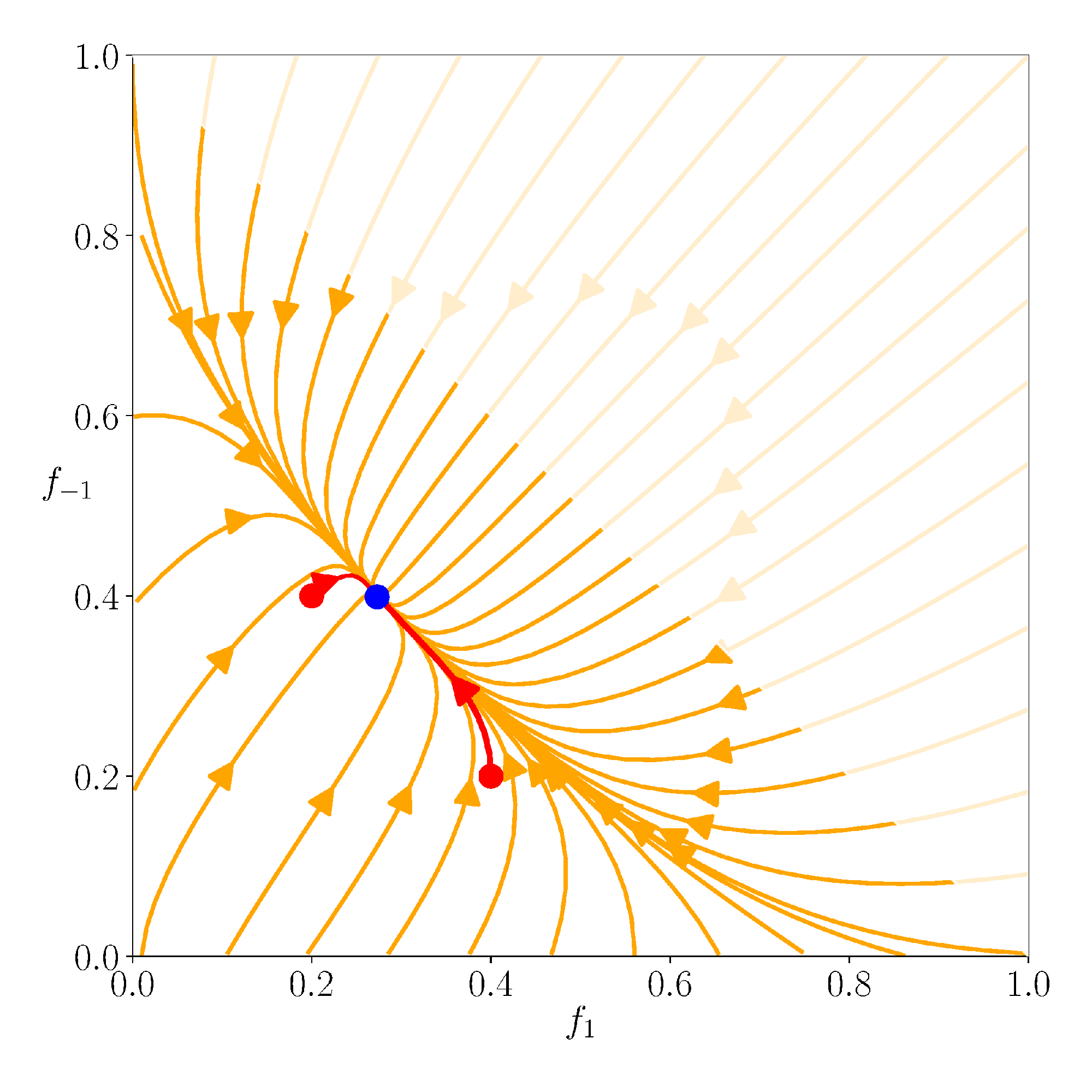}
\includegraphics[scale=0.3]{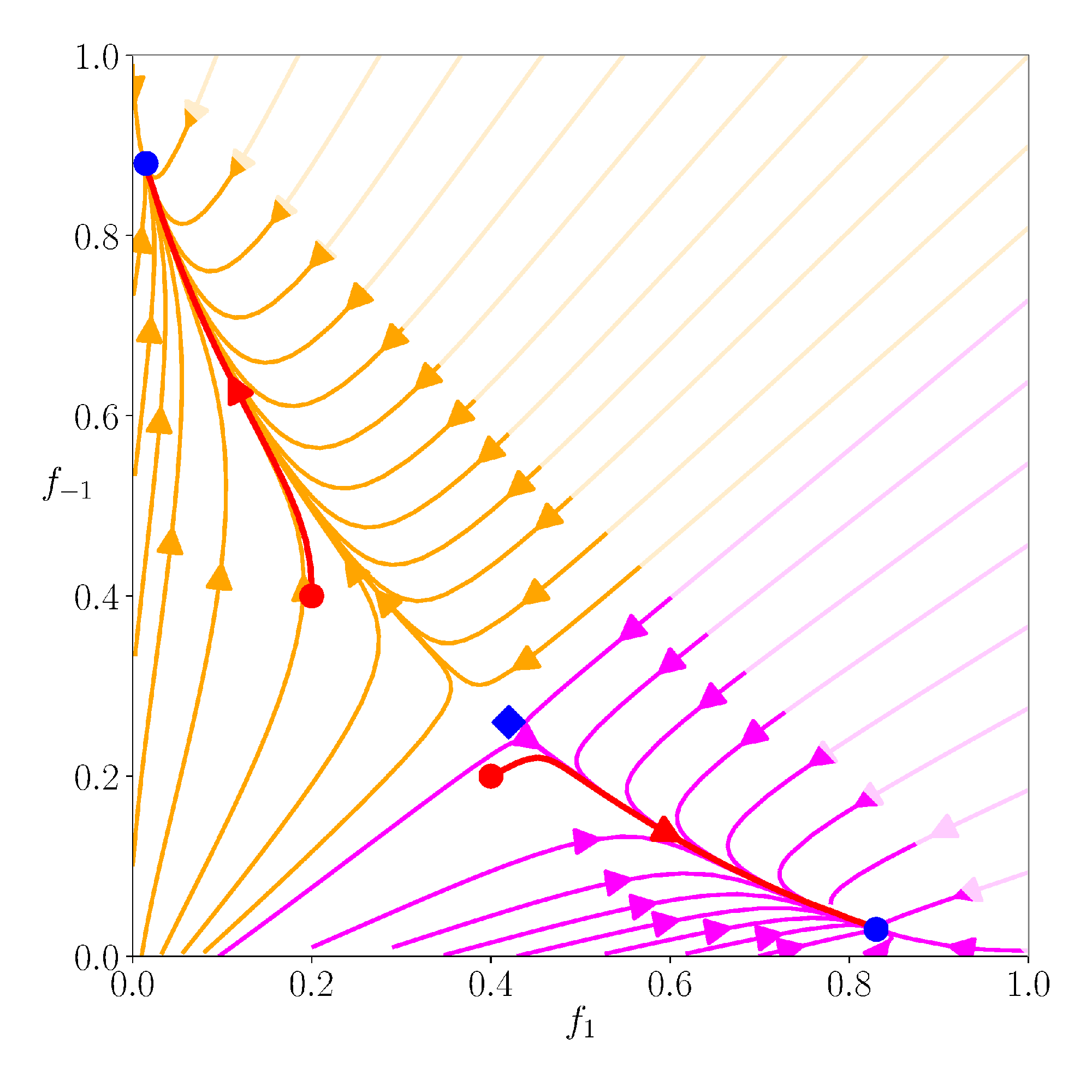} \\
	\mbox{}\hspace{1cm}	  (a) 	\hspace{5.5cm} (b)	 
\caption{Flux diagram in the plane $(f_1,f_{-1})$, 
	for the same parameters used in Fig.~\ref{fig:evolution}. 
	Red points highlight initial conditions and red lines the trajectories 
	toward the stable fixed points (blue circles). The blue square corresponds 
	to a saddle point in the separatrix between the two basins of attraction.
	Notice that only the region $0 \le f_k \le 1$, for $k=-1,0,1$, is accessible.
	}
		\label{fig:diagrams}
\end{figure}

The dashed lines in Fig.~\ref{fig:evolution} were obtained via numerical integration 
of Eqs. (\ref{taxes1}) and (\ref{taxes2}), 
showing a complete agreement with the results of Monte Carlo numerical simulations. 
The influence of initial conditions can be visualized through the flux diagrams in Fig.~\ref{fig:diagrams}, 
where the same two initial conditions of Fig.~\ref{fig:evolution} (red circles)  are plotted, 
together with the fixed points (blue symbols). 
In the left panel, all flux lines converge to a single stable fixed point, 
while in the right panel, there are two basins of attraction, with initial conditions evolving  
towards two distinct stable fixed points (blue circles).

\begin{figure}[h]
	\includegraphics[scale=0.6]{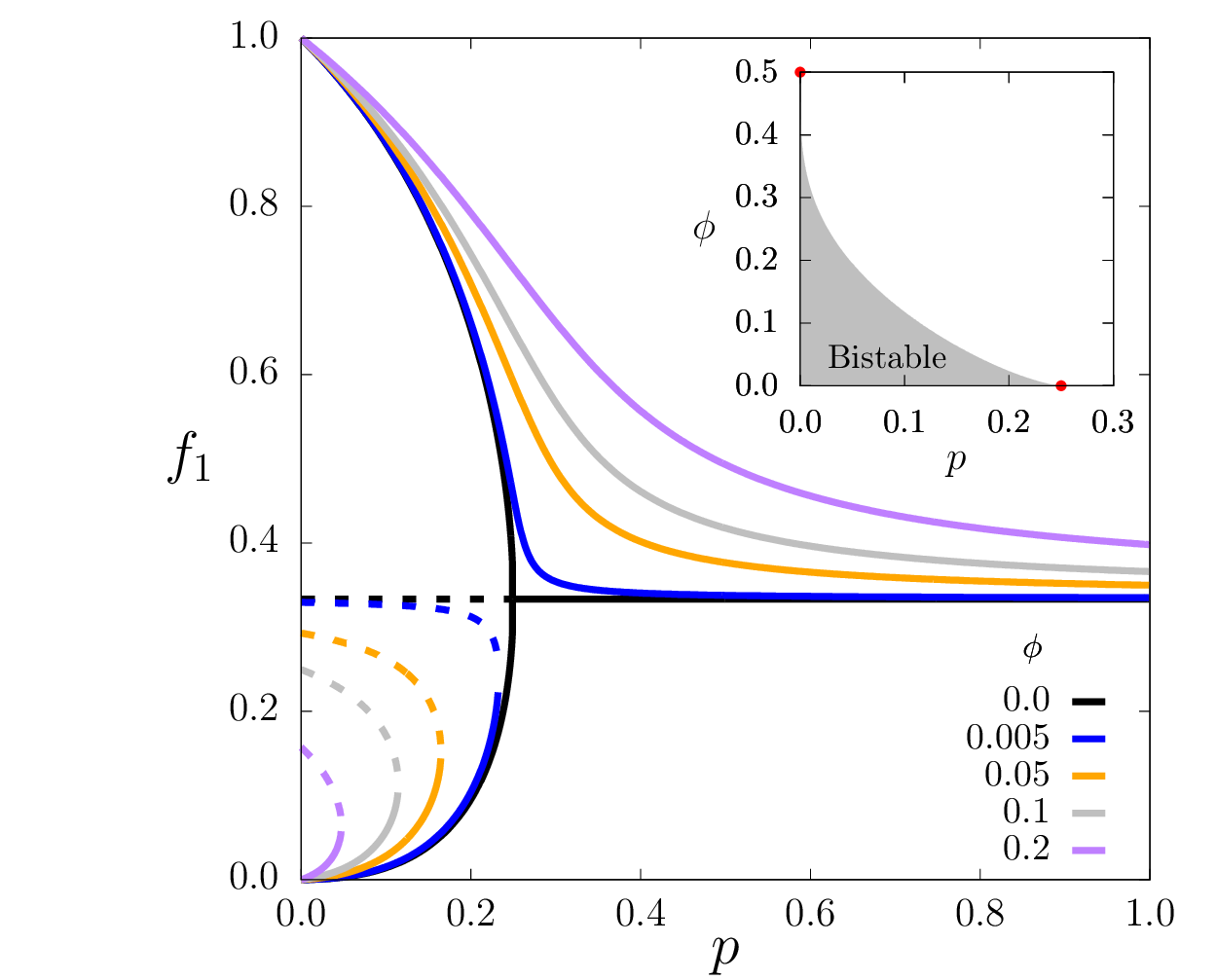}
	\includegraphics[scale=0.6]{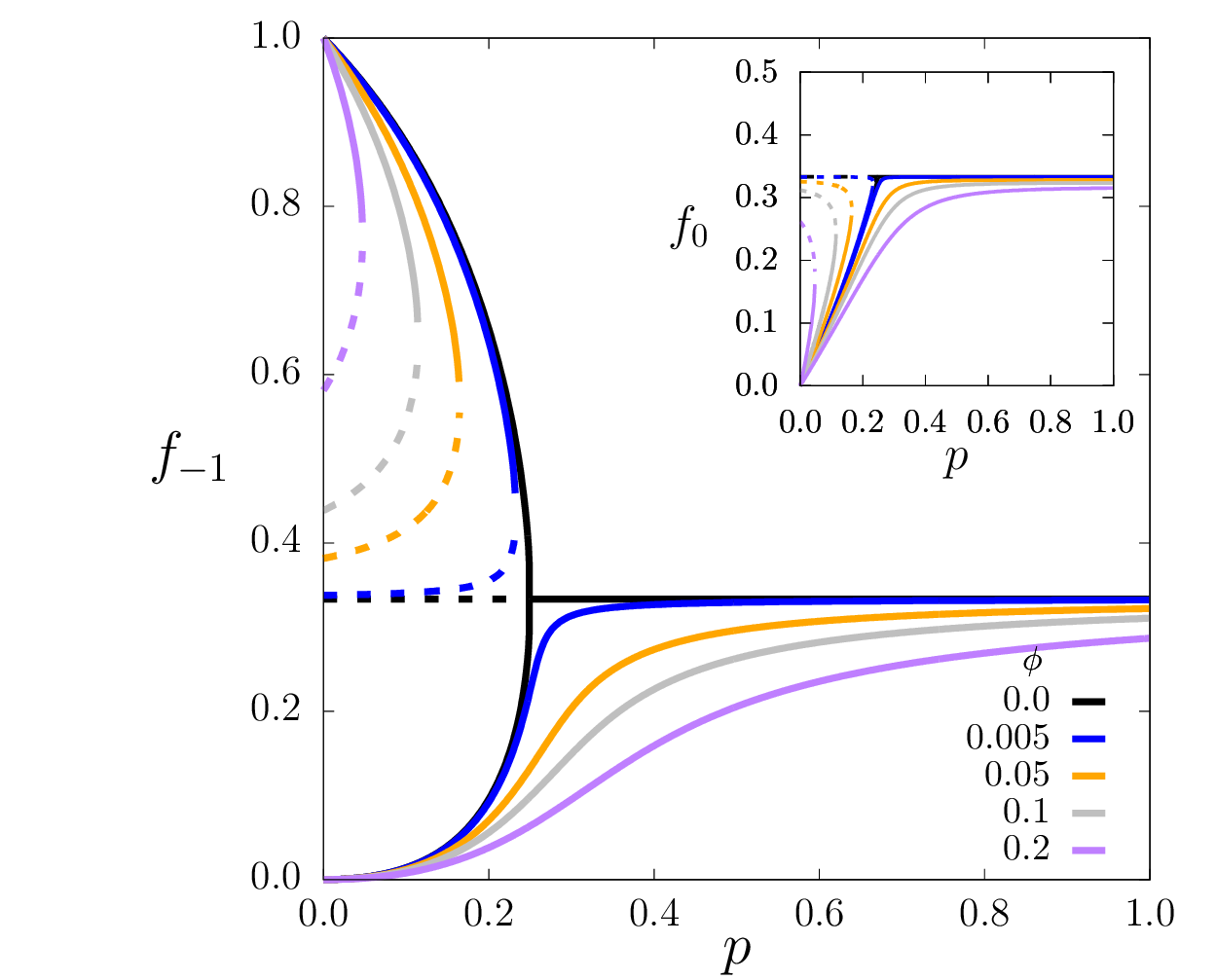}\\
		\mbox{}\hspace{3cm}	  (a) 	\hspace{7cm} (b)	 
	\caption{Bifurcation diagrams of  $f_{1}$ (left)  and $f_{-1}$ (right),  
	as a function of $p$, for fixed values of $\phi$ (indicated in the legend). 
	Solid (dashed) lines correspond to stable (unstable) fixed points. 
	The inset in (a) highlights the region of phase space where there is bistability. 
	The inset in (b) shows corresponding plots for $f_0$.
	}
	\label{fig:bifurcation}
\end{figure}

In order to investigate how the full portrait is altered 
depending on the intensity of the external field, 
we plot in 	Fig.~\ref{fig:bifurcation}  the stability diagrams 
for different fixed values of $\phi$. Solid (dashed)
lines represent stable (physical) solutions and dashed lines unstable ones.
The bistability observed in Figs.~\ref{fig:evolution} and \ref{fig:diagrams} 
is associated to the existence of two branches in these bifurcation diagrams.

For the case where $\phi = 0$, there is a (supercritical) pitchfork bifurcation, 
and two solutions with complementary values of $f_1$ and $f_{-1}$. 
This is in accord with the continuous phase transition observed for the order parameter 
$|\Delta|=|f_{1}- f_{-1}|$, at the critical point $p_c=0.25$. 
Due to the symmetry, $|\Delta|$ vs $p$ produces a single curve. 
For the disordered phase ($\Delta=0$, for $p>p_c$), $f_0=1/3$ and in the 
ordered phase $f_0=p/(1-p)$~\cite{biswas2012disorder} (also see Appendix, and black line in the 
inset of Fig.~\ref{fig:bifurcation}(b)).

Differently, when a small field is switched on,  the pitchfork branches detach. 
(Hence, the continuous transition observed for $\phi=0$ is suppressed, 
because the critical curve splits into two.)   
Despite the imperfect bifurcation that arises for $ 0<\phi <  \phi_c=0.5$, 
the field may still be  unable to produce a winner, depending on the initial conditions. 
Only stronger fields ($\phi \geq \phi_c$, see inset of Fig.~\ref{fig:bifurcation}(a)) 
are capable of imposing a winner independently of the initial configuration, since 
$f_{-1}$ present a single branch above that tipping point. 

Also notice, in the inset of Fig.~\ref{fig:bifurcation}(b), the behavior of the density of 
undecided people $f_0$. In contrast to the case with no field, $f_0$ has two branches 
below the tipping point $p_c(\phi)$. 
Actually this is a consequence of the lack of complementarity 
of $f_1$ and $f_{-1}$. For the states in the lower (upper) branch of $f_0$, 
the number of undecided people diminishes (increases) with increasing field. 

\begin{figure}[h]
	\centering
	\includegraphics[scale=0.6]{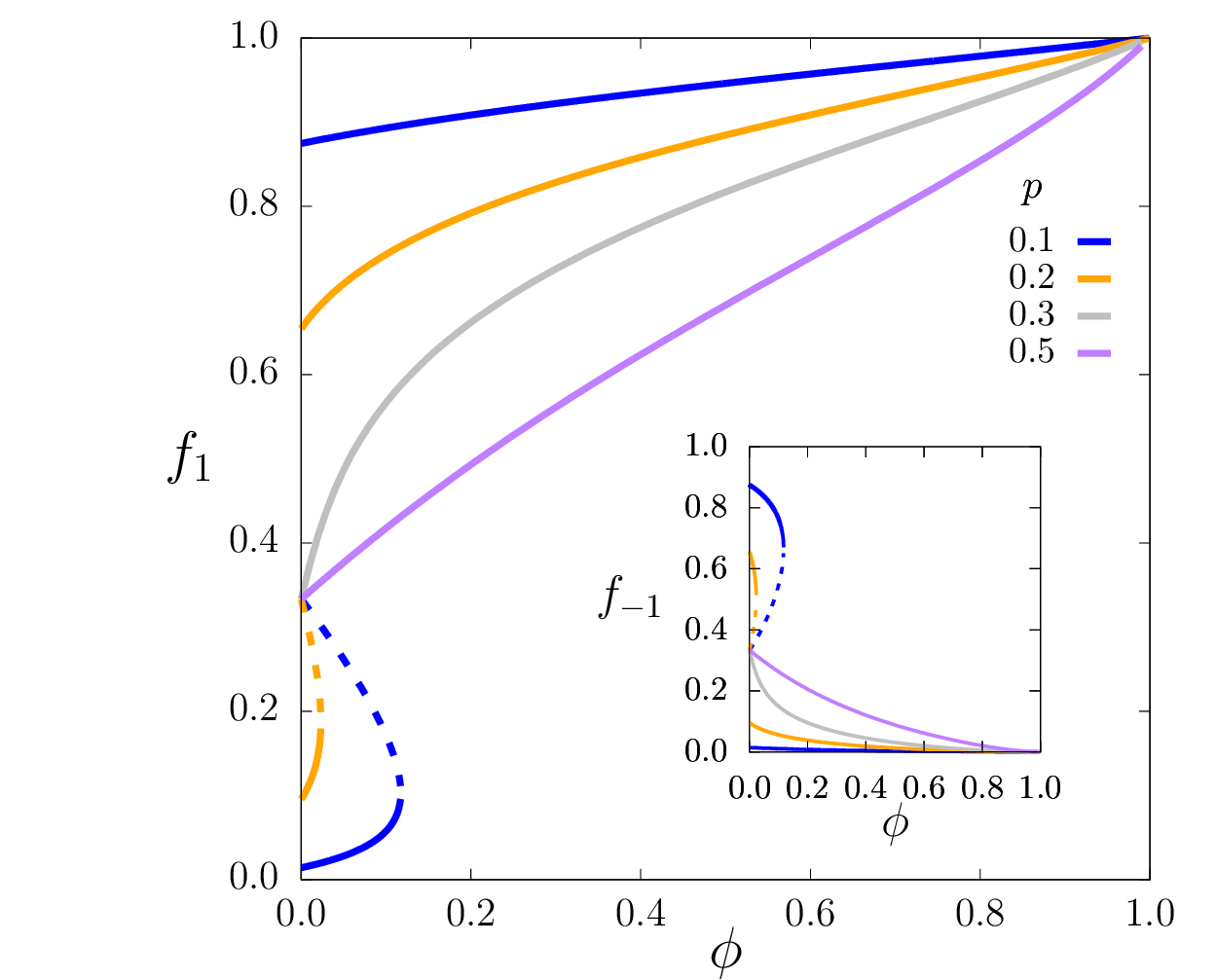}
	\caption{Bifurcation diagrams of $f_1$ as a function of $\phi$, for fixed values of $p$ 
	indicated on the legend ($\omega=1$). 
	The inset shows  the corresponding plots for $f_{-1}$.
		}
	\label{fig:fixed-p}
	\centering  
\end{figure}

It is also worth observing the effect of the field for fixed values of $p$, which means to 
consider a society with a given level of nonconformity. 
In Fig.~\ref{fig:fixed-p}, we plot the bifurcation diagram for $f_1$ 
as a function of $\phi$ ($\omega=1$) for several values of $p$. 
From this point of view, bistability is possible if $p$ is low enough. 
In this case, if the system is in the lower branch, then, 
by increasing the field $\phi$ above the tipping point,  
it will jump to the upper branch. 
Once in the upper branch, 
it cannot come back to the lower one just by reverting the change of the field, 
or even by completely turning off the field, 
except by a strong finite-size fluctuation. 
%~\cite{abramiuk2019independence}.
%
Differently, a large enough proportion of contrarians ($p>0.25$) 
eliminates the lower branch of the steady solution for $f_1$, 
making opinion $o=1$ win, independently of the initial configuration of the system. 
That is, contrarians make the field more effective, by muffling the contribution 
of mutual interactions. 

Let us analyze now the case $0.5\le \omega < 1$, where a contribution of an opposite 
field appears.  
Recall that while $\phi$ controls the sensitivity to any external field, 
$\omega$ is the relative strength of the field favorable to opinion $o=1$. 
Therefore, for $\omega=0.5$, both fields have the same presence in the media, 
while as $\omega$ gradually goes to 1 (or 0),  the field favorable to $o=1$ (or to -1) 
has the strongest impact.

\begin{figure}[h]
 \includegraphics[scale=0.6]{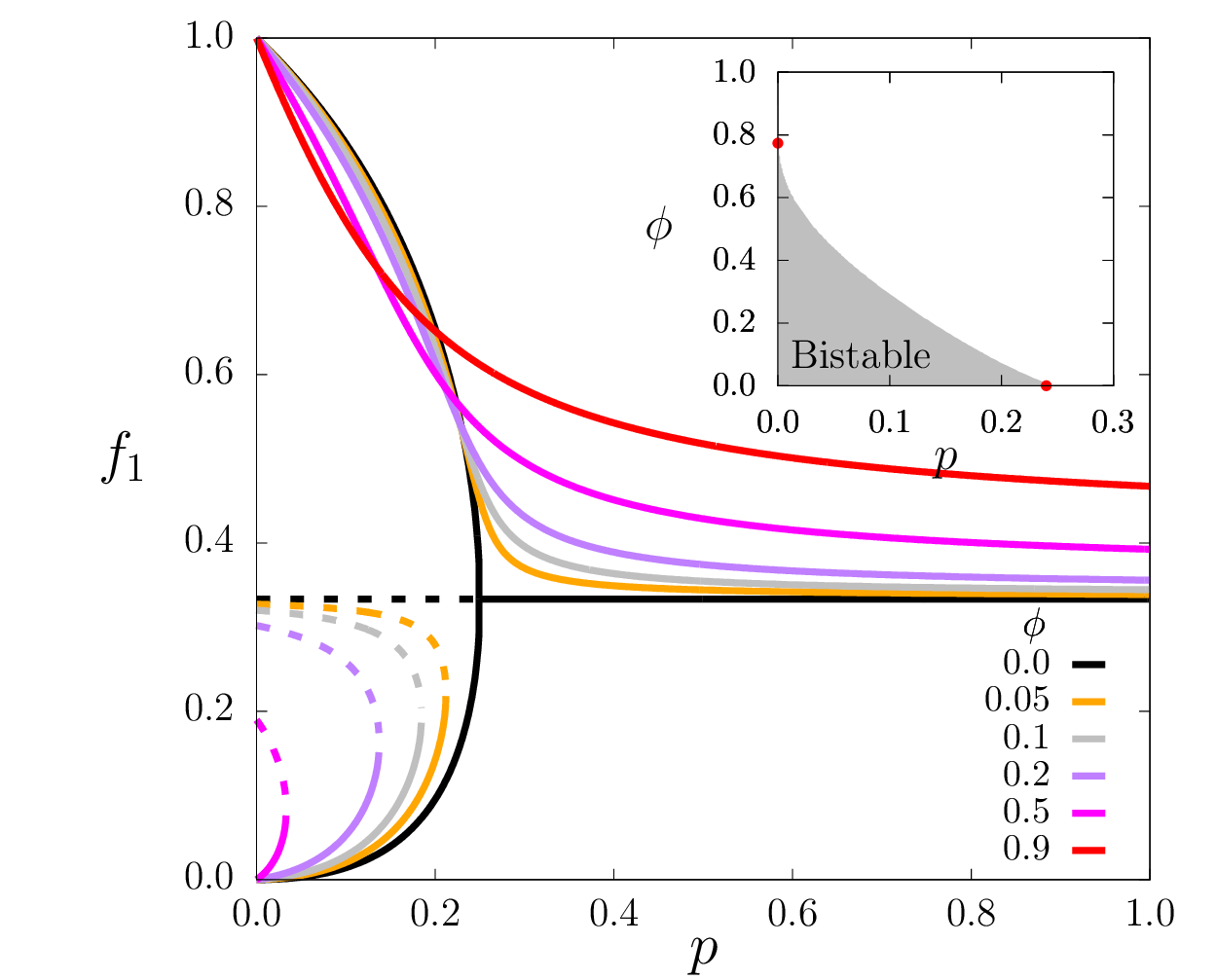}
 \includegraphics[scale=0.16]{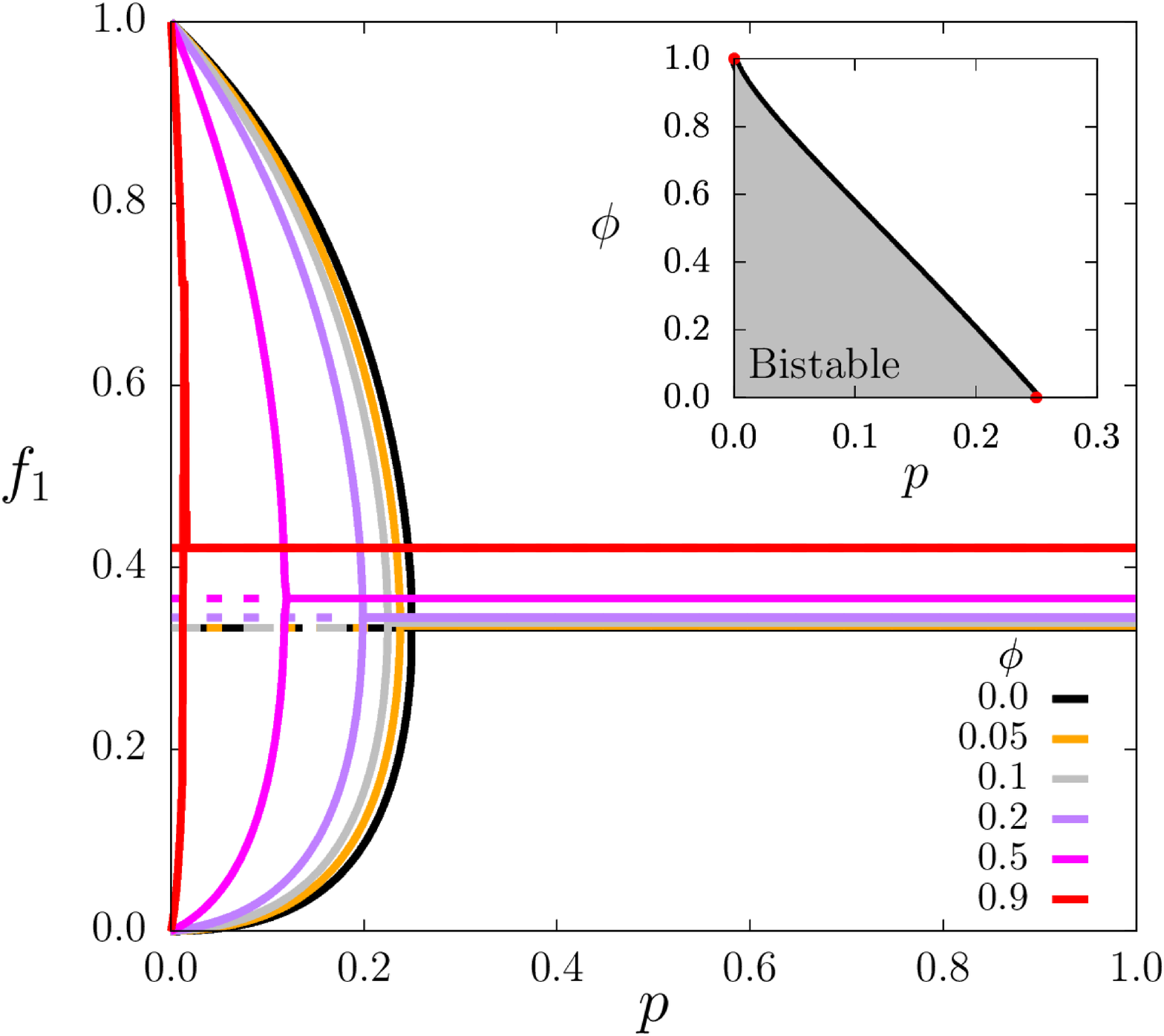}\\
		\mbox{}\hspace{3cm}	  (a) 	\hspace{7cm} (b)	 
	\caption{Bifurcation diagrams showing $f_{1}$ vs. $p$ for different values of $\phi$ 
	(the same used in Fig.~\ref{fig:diagrams}), with  $\omega=0.6$ (a) and $\omega=0.5$ (b).  
	In (a) the inset shows the stability diagrams in the plane $p-\phi$. 
	In the inset (b), the solid line is given by Eq.~(\ref{eq:pc-w05}) and the level 
	of $f_1= (1-f_0)/2$ independent of $p$ in the disordered phase is given 
	by Eq.~(\ref{f0_desordenada}).  
	\label{fig:fixed-w}
	}
\end{figure}

Figure~\ref{fig:fixed-w} shows the behavior of $f_1$ 
as a function of $p$ for several values of $\phi$, 
with $\omega=0.6$  (a) and $\omega=0.5$ (b).  
These plots must be compared to those in Fig.~\ref{fig:diagrams}(a), 
which corresponds to $\omega=1$. 
Notice, that decreasing $\omega$ from 1 to 0.5, 
allows bistability to occur for increasing values of $\phi$ and the 
diagrams tend to a perfect pitchfork, which is attained at $\omega=0.5$ (equal strength of both opposed fields), illustrated in Fig.~\ref{fig:fixed-w}(b).  

For $\omega=0.5$, above the critical value, the system is disordered, in the sense that both extreme opinions are balanced 
($f_1=f_{-1}=(1-f_0)/2$), where the undecided fraction $f_0$ is given by Eq.~(\ref{f0_desordenada}).
The exact expression for $p_c(\phi)$ plotted in the inset of Fig.~\ref{fig:fixed-w}(b) is derived in the Appendix. 
This critical point decreases with $\phi$. 
Hence, in comparison to the absence of fields ($\phi=0$), 
when there are two balanced fields ($\omega=0.5$),
lower values of $p$ (low counter-imitation) 
are enough to promote a disordered state where there is no dominant opinion. 
Therefore, opposed fields have a disorganizing effect. 
Moreover, the fraction of undecided people decreases as $\phi$ increases. 
The continuous phase transition observed in this case, is suppressed if $\omega \neq 0.5$. 

\begin{figure}[h]	
\includegraphics[scale=0.60]{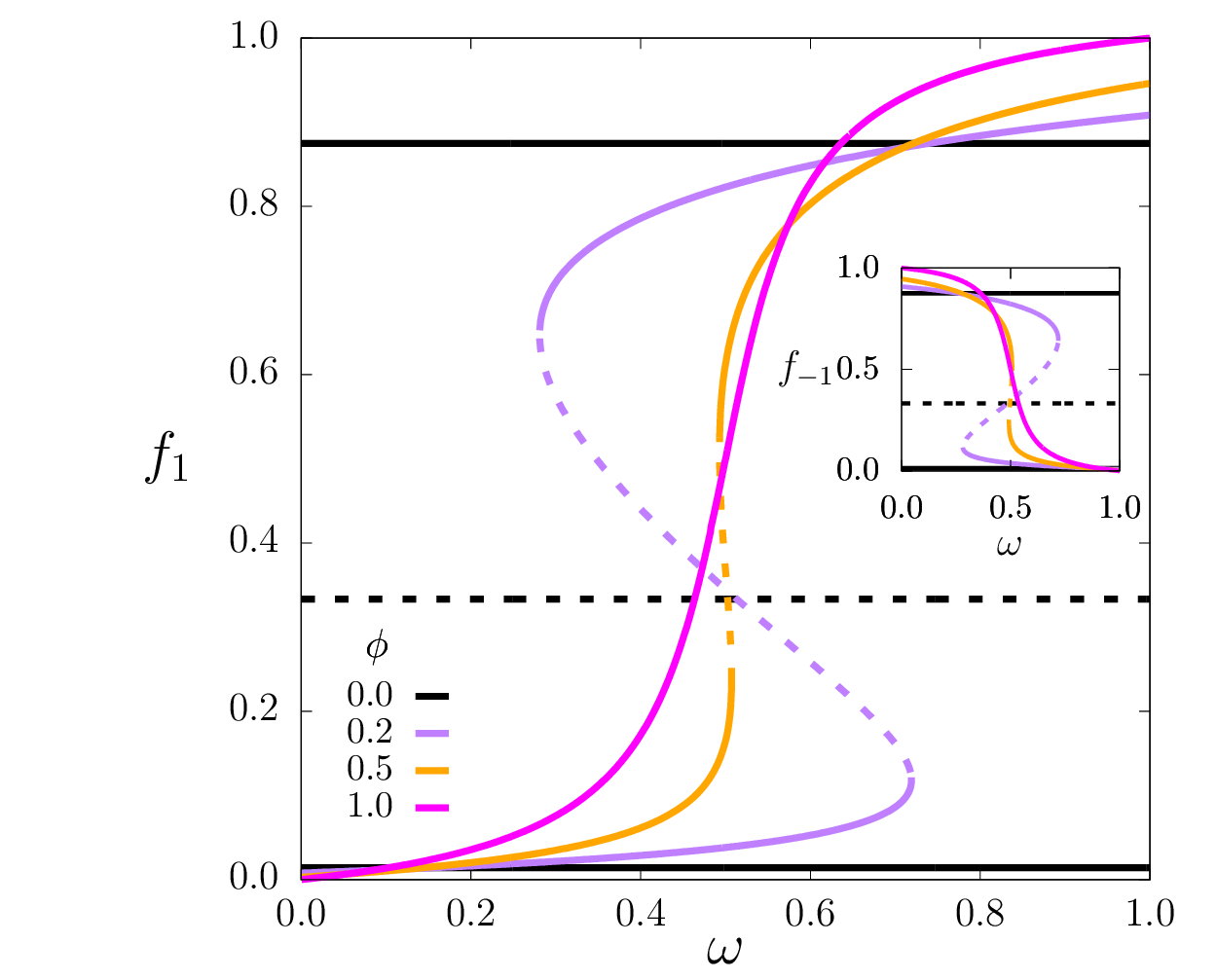}
\includegraphics[scale=0.17]{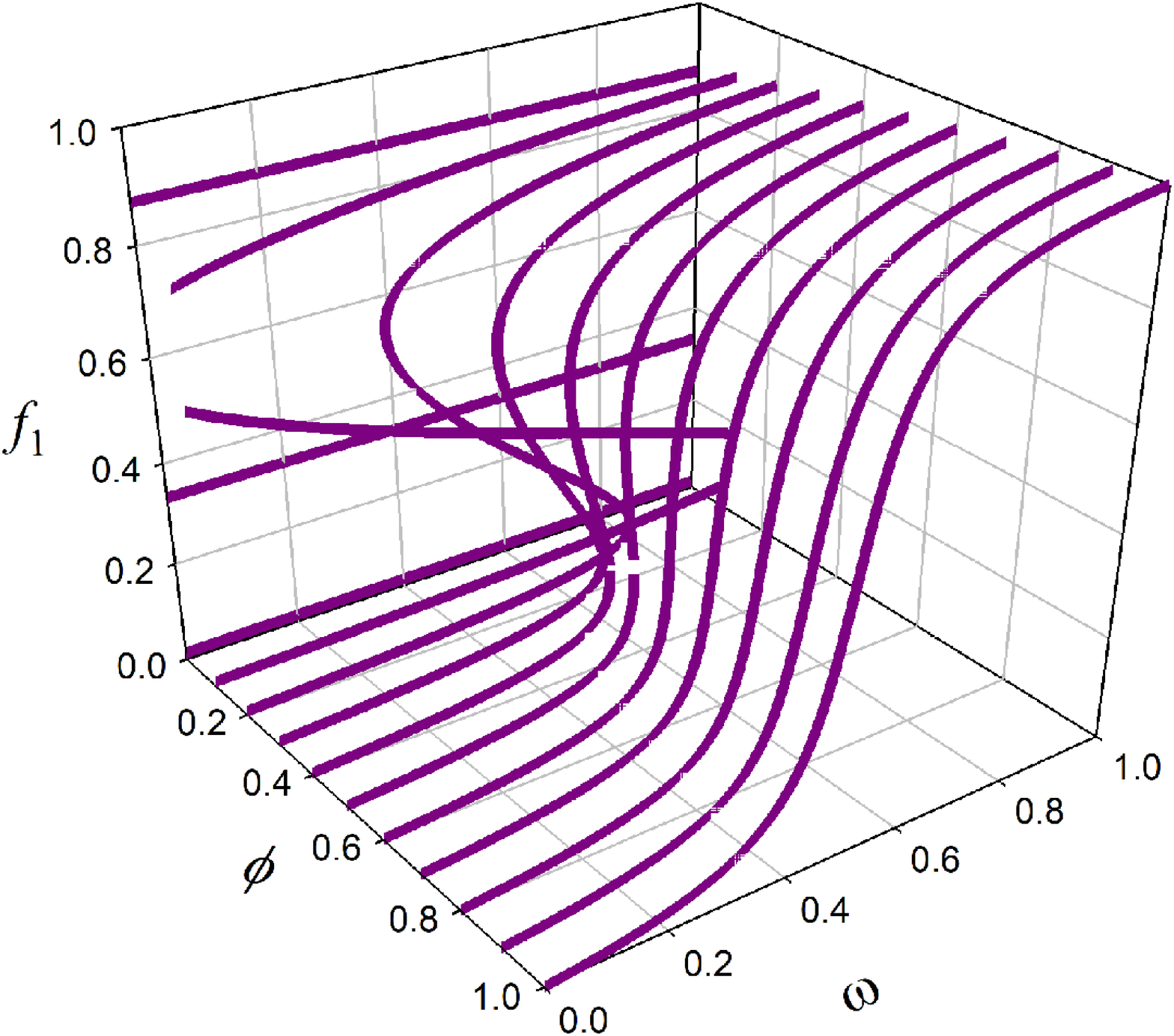}\\
		\mbox{}\hspace{3cm}	  (a) 	\hspace{7cm} (b)	 
\caption{Bifurcation diagram 	of $f_{1}$ vs. $\omega$ 
	for different values of $\phi$ (a), and vs $\phi$ and $\omega$ (b). 
	The change of regime (from bistable to monostable) occurs at $\phi\simeq 0.566$ 
	Disagreement level is $p=0.1$.
	The inset illustrates the symmetry around $\omega=0.5$ under the transformations 
	$\omega \to 1-\omega$ and $f_k \to f_{-k}$.
	}
		\label{fig:www}
\end{figure}

Plots of $f_1$ vs $\omega$ are presented in Fig.~\ref{fig:www}, 
for several values of $\phi$.
Parameter $\omega$ changes the predominance of the fields, being balanced at $\omega=0.5$. 
Then, around this value, the roles of $o=1$ and $o=-1$ are inverted, as shown in 
the inset of Fig.~\ref{fig:www}(a). 
When increasing $\omega$ from 0.5 to 1 (meaning increasing predominance of the field associated to opinion $o=1$), 
the system can eventually jump to a distant fixed point, and recovery is not possible, even if the dominant field is reversed, except when undergoing a large finite-size fluctuation, 
configuring an irreversible transition with hysteresis. 
This is similar to the hysteresis behavior observed by changing $\phi$ with $\omega=1$ in Fig.~\ref{fig:fixed-p}.  A full landscape of $f_1$ vs $\phi$ $\omega$ 
for a fixed level of disagreement is depicted in Fig.~\ref{fig:www}(b).

\section{Final remarks}
\label{sec:remarks}

We investigated the effects of external fields in a kinetic exchange opinion dynamics, 
observing how interpersonal and global external influences drive the collective dynamics. 
In the absence of external fields, when opinions are taken by exchanges between peers, 
there is a pitchfork bifurcation 
at a critical value $p_c=0.25$, signaling a continuous order/disorder transition. 
This kind of diagram also emerges when there are balanced external influences. 
When an external  bias is turned on,  the pitchfork disconnects into two pieces 
as depicted in Fig.~\ref{fig:bifurcation}. 
This has important implications concerning the fate of the collective state when the 
social group is subject to external bias. 
An increase of the field can produce an abrupt change of the winner alternative, 
 which is very difficult to be reversed even if the external field is suppressed, 
as shown in Figs.~\ref{fig:fixed-p} and \ref{fig:www}. 
Therefore the external influence of the media 
can produce  the phenomenon known in nonlinear theory as  {\it catastrophe}.  
It is noteworthy that a similar picture was observed due to  
the presence of zealots~\cite{svenkeson2015reaching}, which somehow act with their 
extremist attitude as a unilateral biased influence, although of internal origin. 

The level of disagreement  in the population plays a role in moderating the effect 
of the field, with its disorganizing role. But opposite fields are disorganizing too.

Concerning indecision, the number of undecided people is reduced by the biased driving, 
since this group is more sensitive to the effect of the field, adopting more easily 
than an opponent the opinion favored by the field. 
Then, an initially large number of people in the undecided state reinforces the effect of 
the field. This is due to the role of the undecided in mediating 
exchanges~\cite{crokidakis2012role,balenzuela2015undecided,vazquez2004ultimate} and also to their larger susceptibility.

Finally let us note that, these results may give insights  to understand more complex situations 
where opinions co-evolve with other dynamics such as epidemics~\cite{pires2018sudden}, 
where the epidemic state can act as a (time variable) field molding opinion dynamics.

%\newpage

\appendix

\section{Derivation of the rate equations}

We obtain the matrix of transition probabilities in which the element $m_{r,s}$ 
is the probability of the change of state $r$ $\rightarrow$ $s$, 
following a simple mean-field approach.
We consider  all possible combinations in Eq.~(\ref{model}), taking into account that 
opinions $o_i,\, o_j \in  \{ -1,0,1 \}$, couplings $\mu_{ij} \in \{ -1,1 \}$, according to Eq.~(\ref{mu}), 
while the effective field can take values in $ \{ -1,0,1 \}$, according to Eq.~(\ref{F}). 
Then, we obtain
\begin{eqnarray*} 
		m_{1,1} &=& f_1\biggl(    f_1(p\phi \omega+1-p)   +f_0(\phi \omega +1-\phi)  +f_{-1}(p+(1-p)\phi \omega) \biggr) \,, \nonumber \\	
		m_{1,0} &=& f_1\biggl(    f_1 p(1-\phi)      +f_0\phi(1- \omega)        +f_{-1} (1-p)(1-\phi)  \biggr) \,, \nonumber \\ 
		m_{1,-1} &=& f_1\biggl(   f_1 p\phi(1-\omega)                           +f_{-1} (1-p)\phi(1-\omega)  \biggr) \,, \nonumber \\ 
		m_{0,1} &=&  f_0\biggl(   f_1(1-p)(\phi \omega+1-\phi)        +f_0 \phi \omega   +f_{-1}p(\phi \omega+1-\phi) \biggr) \,, \nonumber \\	
		m_{0,0} &=&  f_0\biggl(  f_1( p\phi \omega+ (1-p)\phi(1-\omega))  +f_0 (1-\phi)   +f_{-1}( (1-p)\phi \omega + p\phi(1-\omega) ) \biggr) \,, \nonumber \\	
		m_{0,-1} &=&  f_0\biggl(  f_1 p(1-\phi \omega)             +f_0 \phi (1-\omega)   +f_{-1}(1-p)(1- \phi \omega ) \biggr) \,, \nonumber \\	
		m_{-1,1} &=& f_{-1}\biggl(f_1 (1-p) \phi \omega                              +f_{-1}p \phi \omega  \biggr) \,, \nonumber \\	
		m_{-1,0} &=& f_{-1}\biggl(f_1 (1-p)(1- \phi)        +f_0 \phi \omega         +f_{-1}p (1-\phi)  \biggr) \,, \nonumber \\	
		m_{-1,-1} &=& f_{-1}\biggl(f_1 ( p+(1-p)\phi(1-\omega) ) +f_0 (1-\phi \omega)      +f_{-1}(p \phi (1-\omega) +1 - p) \biggr) \,. \nonumber 	
\end{eqnarray*}

Finally, considering that $f_0=1-f_1-f_{-1}$, the equations for the evolution of each density are
\begin{eqnarray}
		\frac{d f_{1}}{d t} &=& m_{0,1}+m_{-1,1} -m_{1,0}-m_{1,-1}  \equiv  G_{1}(f_1,f_{-1}, p, \phi, \omega),  \label{f1} \\
		\frac{d f_{-1}}{d t} &=&  m_{1,-1}+m_{0,-1} -m_{-1,1}-m_{-1,0}   \equiv  G_{-1}(f_1,f_{-1}, p, \phi, \omega).   \label{f-1}  
\end{eqnarray}

\section{Analytical results for  $\omega = 0.5$}

%For $\omega = 0.5$, in Fig.~\ref{fig:fixed-w}(b), we can compute the critical values of $p$ %explicitly. 
%

The  usual  order parameter is defined as $O=  |f_1 - f_{-1}| $. The evolution of 
$\Delta = f_1 - f_{-1}$, from Eqs.~(\ref{f1}) and (\ref{f-1}), when $\omega=0.5$, 
can be cast in the form 
\begin{equation}
\frac{d \Delta}{dt}= \biggl( \frac{d f_1}{dt} - \frac{d f_{-1}}{dt} \biggr)
= a \, \Delta \,.
\end{equation}
where $a=f_0 (1-p)(1-\phi)-p$. 

In the steady state, it must be either $\Delta=0$ (meaning a disordered state,
where opposite opinions balance) or $a=0$ (meaning unbalance). 
Therefore, in the ordered phase, the fraction of undecided people is
\begin{equation}\label{f0_ordenada}
f_0 = \frac{p}{(1-p)(1-\phi)}.
\end{equation}

In the disordered phase, $\Delta =0$. Thus, setting $f_{-1} = f_{1}$ in the expression 
$ df_0/dt  = 0$, we obtain 
\begin{equation}\label{f0_desordenada}
f_0 = \frac{2(1-\phi)-\sqrt{1-\phi^2}}{3- 5\phi},  
\end{equation}
after discarding the unphysical solution. 
(At $\phi=3/5$, this expression must be substituted by its limit value.) 
Notice that,  $f_0$ is independent of $p$.
It decreases with $\phi$ from 1/3 (when $\phi=0$) to 0 when ($\phi=1$).

\begin{figure}[h]	
\centering
\includegraphics[scale=0.50]{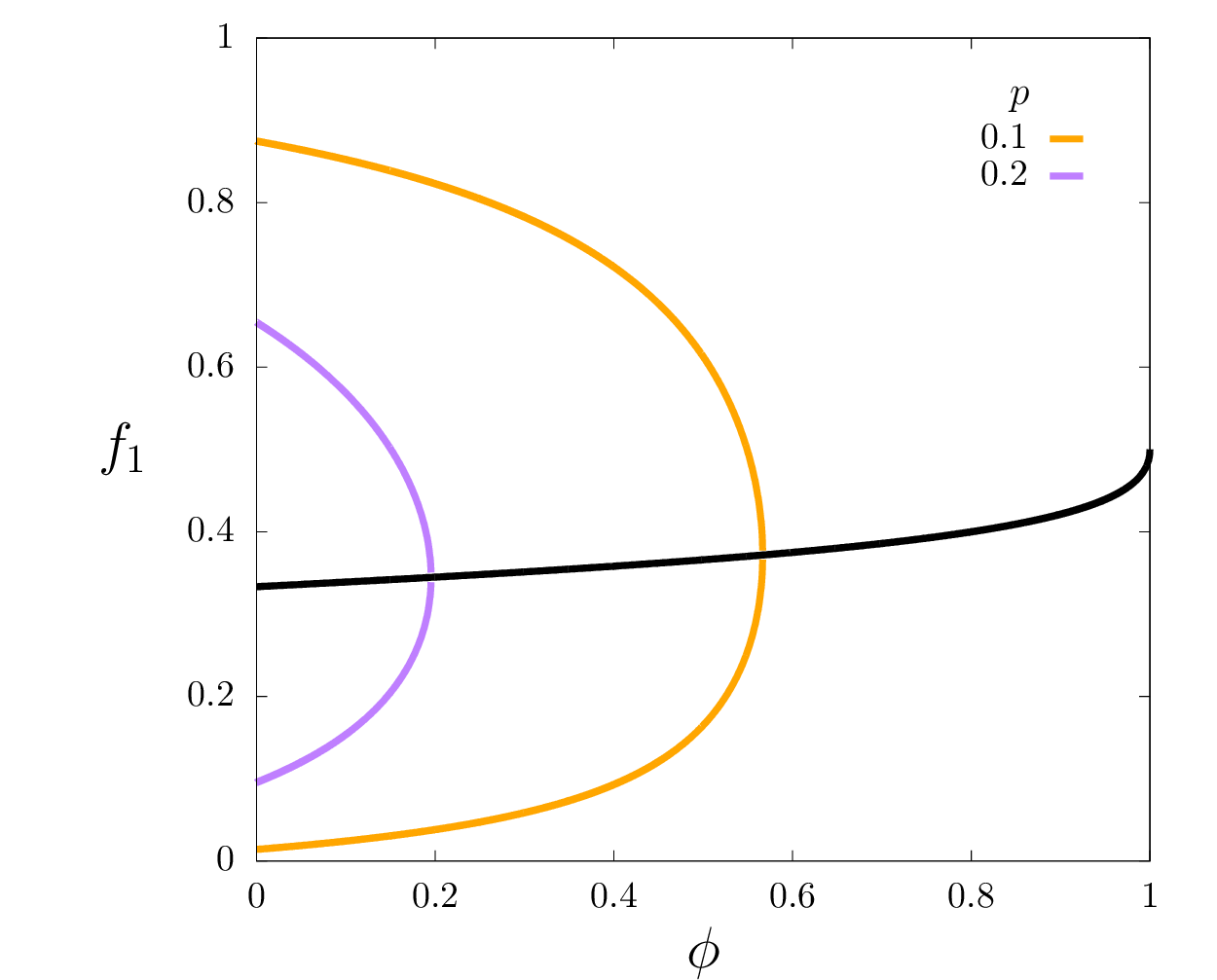}
\caption{Fraction $f_{1}$ vs. $\phi$, for fixed values of $p$, using the analytical procedure 
described in the text. In the disordered phase, the dark line highlights that the result 
is independent of $p$, $f_1=f_{-1}=(1-f_0)/2$. 
}
		\label{fig:f1vsphi}
\end{figure}

The expressions given by Eqs. (\ref{f0_ordenada}) and (\ref{f0_desordenada}) 
must coincide at the critical point, thus yielding
\begin{equation} \label{eq:pc-w05}
p_c = \frac{1-\phi}{3-\phi +\sqrt{(1+\phi)/(1-\phi)}   } \,.
\end{equation}
This curve is plotted in the inset of Fig.~\ref{fig:fixed-w}. 
For $\phi=0$,  the known value $f_0 = 1/3$ and $p_c= 1/4$ are recovered. 
If $\phi=1$, then $p_c=0$.

Finally, we can obtain $f_1$ for both phases. 
In the disordered phase, simply $f_1= (1-f_0)/2$, 
where $f_0$ is given by Eq.~(\ref{f0_desordenada}). 
For the ordered phase, one can solve Eq.~(\ref{taxes1}) analytically 
using $f_0$ from Eq.~(\ref{f0_ordenada}). 
The behavior of $f_1$ vs $\phi$ is illustrated in Fig.~\ref{fig:f1vsphi}. 

It is noteworthy that there are two stable ordered solutions of $f_1$. 
Meanwhile, $f_{-1}$ has the same solutions, assuming complementary values. 
Then, we obtain $|\Delta | \sim |p-p_c|^\beta$ where $\beta=1/2$, 
the typical mean field exponent for continuous phase transition\cite{stanley1971phase}.

\newpage

\bibliographystyle{plain}
%\bibliographystyle{unsrt}
%\bibliography{bibliography}

\end{document}